\documentclass[preprint,12pt,review,3p,sort&compress]{elsarticle}
\usepackage{graphicx}
\usepackage{dcolumn}
\usepackage{bm}
\usepackage{array}
\usepackage{float}

\newcolumntype{C}[1]{>{\centering}m{#1}}
\usepackage{amsmath} 
\usepackage{fancyhdr}
\usepackage{textcomp}

\journal{Carbon}

\begin{document}

\begin{frontmatter}

\title{Tribochemistry of graphene on iron and its possible role in lubrication of steel}

\author[label1]{Paolo Restuccia}
\address[label1]{Dipartimento di Scienze Fisiche, Informatiche e Matematiche, Universit\`a di Modena e Reggio Emilia, Via Campi 213/A, 41125 Modena, Italy}

\author[label1,label2]{M. C. Righi\corref{cor1}}
\address[label2]{CNR-Institute of Nanoscience, S3 Center, Via Campi 213/A, 41125 Modena, Italy}
\cortext[cor1]{Corresponding author}
\ead{mcrighi@unimore.it}

\begin{abstract}
Recent tribological experiments revealed that graphene is able to lubricate 
macroscale steel-on-steel sliding contacts very effectively both in dry and humid conditions.
This effect has been attributed to a \emph{mechanical action} of graphene related
to its load-carrying capacity. Here we provide further insight into the functionality of graphene as 
lubricant by analysing its \emph{tribochemical action}.
By means of first principles calculations we show that graphene binds strongly to native
iron surfaces highly reducing their surface energy. Thanks to a passivating effect,
the metal surfaces coated by graphene become almost inert and 
present very low adhesion and shear strength when mated in a sliding contact.
We generalize the result by establishing a connection between the tribological and the electronic properties of interfaces,
which is relevant to understand the fundamental nature of frictional forces.
\end{abstract}


\end{frontmatter}

\section{Introduction}

In recent years the search for novel lubricant materials and coatings 
has gained increasing importance to face the
massive economic and environmental costs related to friction and wear.
Furthermore, the development of miniaturized devices with high surface-to-volume ratio, such as micro- and nano-electromechanical systems,
quests for new solutions for tribological problems like stiction that seriously undermine 
their functionality. Graphene is regarded as a new emerging lubricant with great potentialities in this context.\cite{Erdemir_review,TysoeCutEdge}
It can be used as solid, atomic-thick coating or as colloidal liquid lubricant. 
Its high strength, chemical stability and easy shear capability 
make it very appealing for a wide range of tribological applications, as revealed by the increasing number
of experimental\cite{FilleterPRL,LeeNanotech,LiPSSB,LeeScience,Kwang-SeopACSN,LinSCT,FengACSN,DengNM,MarchettoTL,BermanAPL,MarchettoF,OtaRSCA,LiangCarbon} and theoretical studies\cite{GuoPRB,BonelliEPJB,LiangNanotech,XuCarbon,LiuCarbon,SmolyanitskyPRB,graphene_reguclelia,mio_PRB_marco,
HodPRB,HodJPCL,CiraciPRB,Moseler,CiraciSpringer} on graphene friction.
Recent experimental findings, in particular, revealed that graphene posses a great potential 
as solid lubricant not only for nano-scale, but also for macro-scale applications.
It has, in fact, been shown that a small amount of graphene-containing ethanol solution
is able to decrease the wear of steel surfaces by four orders of magnitudes
and their coefficient of friction (COF) by a factor of six.\cite{Erdemir_grafhum}  
The lifetime of the slippery regime decreased with the applied load\cite{Erdemir_grafdry}  
as also observed in other previous works.\cite{MarchettoTL,Kwang-SeopACSN,LinSCT,ShinC,WählischW}\\
A possible explanation for these results has been provided by Klements \emph{et al.},\cite{Moseler}
who combined classical molecular dynamics simulations and AFM experiments on
graphene-covered Pt(111), and showed that the ability of graphene to reduce the COF
rests on its ability to increase the load-carrying capacity of the surface. This reduces the
penetration depth of the tip and consequently the friction and wear.
Once the graphene has ruptured, the tribological behavior of the bare metal substrate is recovered.
Therefore the authors conclude that graphene can be an excellent coating for low friction and wear as far as it is not damaged.
\par In this paper we add a new piece of information for understanding  the lubricating properties of graphene
at an atomistic level
by analysing the role played by the surface chemistry.
We consider iron, which is different from
steel, but it may constitute a suitable model for
the native metal surfaces exposed during scratching.\cite{Philippon2011}
We show that graphene chemically adsorbs on native iron surfaces
and its adhesion to the metal substrate is enhanced by the presence of carbon dangling bonds. 
By analysing the interaction between iron surfaces covered by graphene, we 
observe that the atomic-thick carbon layer is able to screen the metal-metal interaction
and dramatically reduce the interfacial adhesion and shear strength. 
Therefore we propose that the lubricating properties of graphene reside in 
its ability to passivate the metal surfaces and suggest an alternative explanation 
for the detrimental role of load, which is of peeling-off graphene from the surface,
thus reducing the graphene coverage. Finally, we analyse the electronic charge displacements
occurring upon interface formation and show that an important connection between
the interfacial electronic and tribological properties can be established,
which can open the way to further investigations on the fundamental nature of the frictional forces.

\section{Method}
We perform spin-polarized density functional theory (DFT) calculations, where the ionic species are described by pseudopotentials and the electronic wave-functions expanded in plane waves.\cite{pw}
The pseudopotential used for iron contains nonlinear core corrections. 
On the basis of test calculations of structural properties of iron bulk and isolated graphene, a kinetic energy cutoff of 30 Ry (240 Ry) is used to truncate the plane-wave expansion of the electronic wave functions (charge density). The Brillouin zone samplings of the supercells used to model graphene layer (GL) and graphene ribbons (GRs) on Fe(110) (described below) are realized by means of $2 \times 1$ and $1 \times 2$  Monkhorst-Pack grids,\cite{monkhorst} respectively.
The exchange correlation functional is described by the generalized gradient approximations (GGA) calculated with the Perdew-Burke-Ernzerhof (PBE) parameterization.\cite{gga-pbe}  We take into account the van der Waals (vdW) interactions, since they are important to realistically  describe the graphene-graphene interaction. To this aim we use the DFT-D semi-empirical approach proposed by Grimme, with a scaling factor $s_6 = 0.75$.\cite{grimme} The results obtain within the DFT-D scheme are compared with those obtained in the local density (LDA) and PBE approximations.  
\par The interfaces are constructed by mating two iron slabs (partially) covered by graphene within the same supercell adopted for surface calculations, this allows to compare total energies. The large number of electronic states involved in the spin-polarized DFT calculations of iron interfaces
imposes the use of a small number of metal layers in order to make the calculations computationally affordable: 
we use four iron layers to simulate interfaces and two layers for surfaces. The slab bottom layer is held rigid during the relaxation processes.

We examine the effects of such approximation on the surface energy, $\gamma$, of iron.
The analysis reported in this paper is, in fact, mainly based on the 
calculation of surface energy differences.
The value of $\gamma$ that we obtain using a slab two-layers thick ($2.27 \; \text{J}/\text{m}^2$ ) is very similar to that
obtained using three- and six-layers thick slabs  ($2.28 \; \text{J}/\text{m}^2$ in both the cases) and these values are in good agreement both with the experimental data ($2.41 \; \text{J}/\text{m}^2$) \cite{TysonSurfEn} and the result of previous calculations ($2.46 \; \text{J}/\text{m}^2$). \cite{VitosSurfEn} 
\par We investigate the effects of graphene coating on tribological properties of iron interfaces. The first property that we consider is the work of separation, $W_{sep}$, defined as the energy per unit area required to separate two surfaces from contact to an ideally infinite separation.\cite{BatyrevPRB} In other words, $W_{sep}$ corresponds to the difference between the formation energy of two isolated surfaces and the formation energy of an interface: $W_{sep} = \gamma_1 + \gamma_2 - \gamma_{12}$. 
By definition, it is assumed that the separated slabs have the same composition of the two slabs joint together to form the interface, therefore the work of separation can be calculated as a difference between total energies: $W_{sep} = (E_1 + E_2 - E_{12})/A$, where A is the supercell in-plane size, $E_{12}$ is the total energy of the supercell containing two slabs in contact and $E_1$ ($E_2$) is the total energy of the same supercell containing only the upper (lower) slab (both the two- and one-slab systems are considered in their optimized configuration, i.e., a relaxation process is carried out). The thickness of the vacuum region present in the supercell adopted in our calculation (about 24 {\r{A} (17  {\r{A}) in the case of one- (two-) slab system) is enough to isolate the system from its periodic replicas.
We calculate $W_{sep}$ for different relative lateral positions of the two surfaces in contact. At each location the structural relaxation is carried out by keeping fixed the bottom layer of the lower slab and optimizing all the other degrees of freedom except for the ($x, y$) coordinates of the topmost layer of the upper slab. In this way, the distance between the two surfaces can reach its equilibrium value, $z_{eq}$, at each fixed lateral position. By exploiting the system symmetries and interpolating, we construct the potential energy surface (PES), $W_{sep}(x,y,z_{eq})$, which describes the variation of the work of separation as a function of surface relative lateral position. The absolute minimum of the PES is register as the
reference value for the work of separation of the considered interface.
\par The second interfacial quantity that we calculate is the shear strength, i.e., the maximum resistance to sliding
of the interface. It is obtained from the derivative of  the PES profile.\cite{mio_langmuir} We consider the PES profile along 
the minimum energy path (MEP), which is the path with the highest statistical weight that connects the PES minima passing trough saddle points. 
The lateral force per unit area experienced by the surface during its displacement, $r$, along the MEP is obtained
as $\tau_{MEP} = -\frac{\partial}{\partial r} W_{sep}(r)$. We register the most negative value of the periodic function $\tau_{MEP}$ as the shear strength of the interface under consideration.

\section{Results and Discussion}

\subsection{Graphene adsorption on native iron}
We start by considering the adsorption of a graphene layer on the (110) surface of iron, which is the most stable surface for this material. The geometry of the Fe(110) surface consists in a distorted hexagonal symmetry, where the nearest neighbor distance among the surface Fe atoms (2.48 \r{A}) is quite close to the graphene lattice constant (2.46 \r{A}). We consider an antiparallel orientation of the graphene layer with respect to the Fe(110) lattice due to the small mismatch present between the iron ($3 \times 5$)
and graphene ($2 \times 8 $) superlattices (about 2.6\% along the [001] direction, where it is maximum). 
The adsorbed graphene sheet maintains a planar structure and adsorbs at an equilibrium distance $d = 2.11 \; \text{\r{A}}$ from the surface (Fig.~\ref{model}a). 
An average distance $\bar{d} = 2.32 \; \text{\r{A}}$ 
has been reported for the corrugated 6 $\times$ 18  superstructure observed in case of parallel alignment 
of the graphene and Fe(110) lattices.\cite{grafonfe_6x18} According to our calculations,
a perpendicular orientation of the two superimposed lattices reduces the interfacial commensurability 
and allows graphene to retain its planar structure. 
The DFT-D binding energy for this configuration is $E_{ad} = -0.89 \; \text{J}/\text{m}^2 = -149 \; \text{meV}/\text{C atom}$.
Such value indicates a relatively strong binging, in agreement with XPS and NEXAFS observations that show that the electronic structure
of graphene adsorbed on iron is significantly disturbed by the substrate.\cite{grafonfe_6x18}\\

\begin{figure}[H]
 \begin{center}
\includegraphics[width=0.7\linewidth]{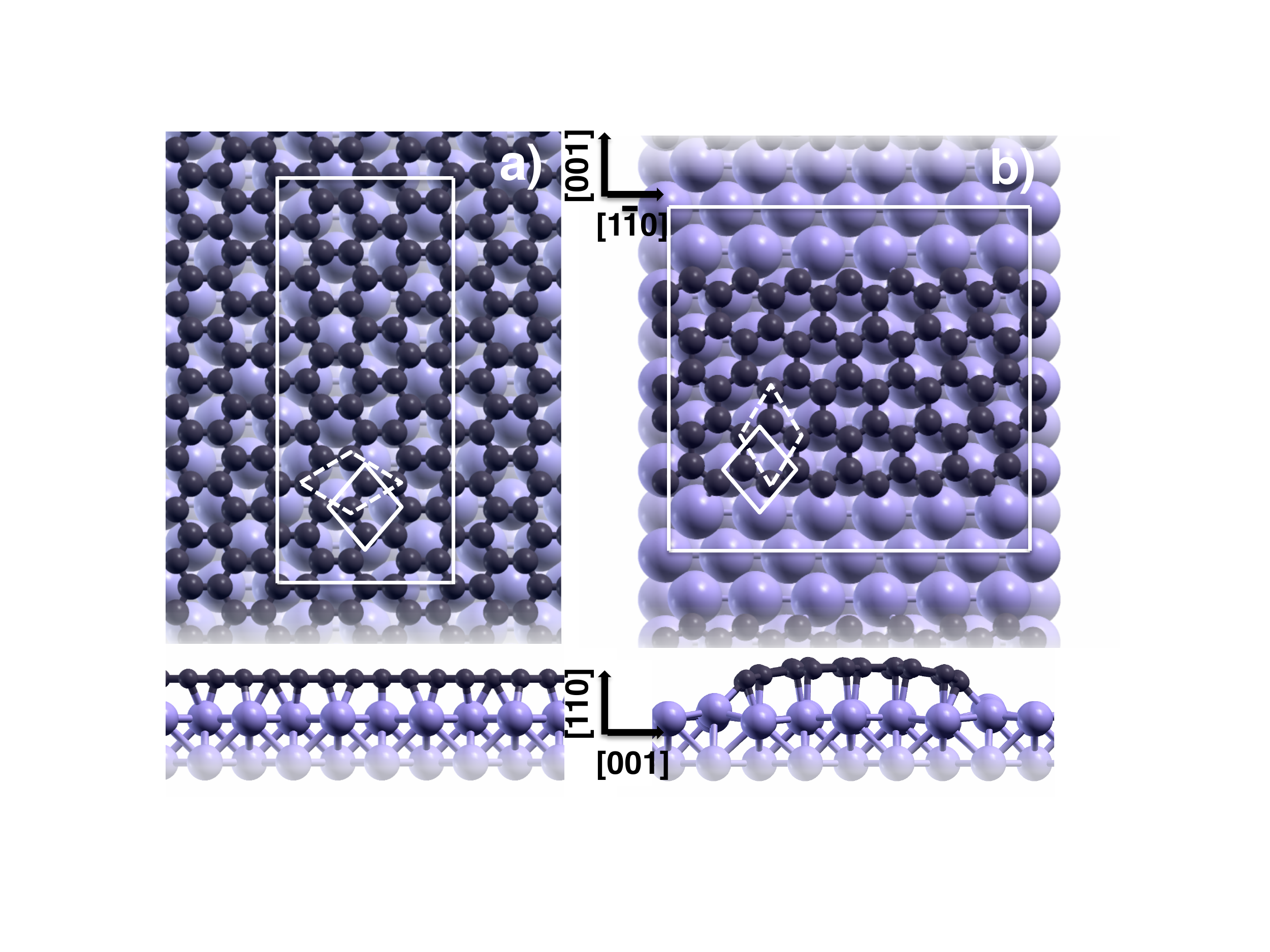}
  \caption{Top and lateral views of the optimized structures of a graphene layer (a)
 and graphene ribbon (b) adsorbed on the Fe(110) surface. The ($3 \times 5$) (a) and ($6 \times 4$) (b)
 two-dimensional supercells used in the 
 calculations are marked in white along with the unit cells of the iron surface
 (continuous line) and graphene (dashed line). }  　\label{model}
\end{center}
\end{figure}

We are not aware of any previous calculation of graphene adsorption energy on iron.
Graphene adsorption has been studied on Co, Ni and Pd and values of $d$ and $E_{ad}$ similar to those reported above have been obtained.\cite{grafonmetals,grafonNi,grafonmetals_review}
The analysis of the band structure of graphene on these substrates reveals a strong perturbation 
due to the hybridization of the $p_z$ orbitals of graphene with the partially occupied $d$ states of the metal.
The corresponding bands acquire a mixed graphene-metal character. This demonstrates that graphene is chemisorbed
on these metals. In contrast, if graphene is adsorbed on metals with the $d$ states fully
occupied, as Al, Cu, Ag, Au, its electronic bands can be clearly identified
and the adsorption distance and energy assume typical values of physisorption
($d \simeq 3.3 \; \text{\r{A}}$, $E_{ad} \simeq -40 \; \text{meV}/\text{C atom}$).\cite{grafonmetals}
Iron belongs, thus, to the first group of metal subtrates on top of which graphene chemically adsorbs.\\
In addition to a complete graphene monolayer, we consider graphene ribbons on iron.
Edge passivation by hydrogen is not taken into account to resemble the
pristine graphene flakes used in the experiments.\cite{Erdemir_grafhum,Erdemir_grafdry}  Zigzag ribbons with the same orientation of the Fe(110) surface 
are simulated using a $6 \times 4$ cell (in terms of the Fe(110) unit cell) and three different ribbon widths are considered to mimic different coverages. The optimized adsorption structure, reported in Fig.~\ref{model}b,
presents the carbon dangling bonds attached to the iron surface. 
The calculated adhesion energy for the ribbon of  Fig.~\ref{model}b, which covers 45\% of the surface, is 63\% higher in absolute value than that of an infinite layer adsorbed on the same surface. 
As shown in the supplemental information (SI), the adhesion energy presents a small variation
as a function of the ribbon width, indicating that
the adhesion of ribbons is dominated by the edge binding to the substrate.
The comparative analysis among the energetics obtained within PBE, PBE+vdW and LDA, reported as SI,
reveals that the PBE (PBE+vdW) approximation is the most appropriate to describe clean (graphene-covered) iron surfaces.

\subsection{Effects of graphene coating on interfacial adhesion and shear strength}
Having characterized graphene adsorption on the surface,
we consider the effects of graphene on interfacial properties.
The interface is modelled by mating two optimized surfaces and then relaxing the whole structure.
The optimized structure obtained for the clean iron interface is shown in Fig.~\ref{interf}a, where interfacial properties
calculated within the PBE approximation are also reported.  The most favorable relative lateral position
of the two mated (110) surfaces turned out to be that corresponding to the stacking position assumed by subsequent (110) planes in bulk iron and the equilibrium distance, $z_{eq}$,
reached by the Fe interfacial layers (indicated by red arrows) at the end of the relaxation process 
corresponds to the typical separation of the bulk atomic planes along the [110] direction. 
The calculated work of separation is equal to twice
the surface energy, as expected from the definition of $W_{sep}$.
These results confirm the validity of both our model for solid interfaces and the approach adopted 
to derive the work of separation from first principles.\\

\begin{figure}[H]
\begin{center}
\includegraphics[width=0.7\linewidth]{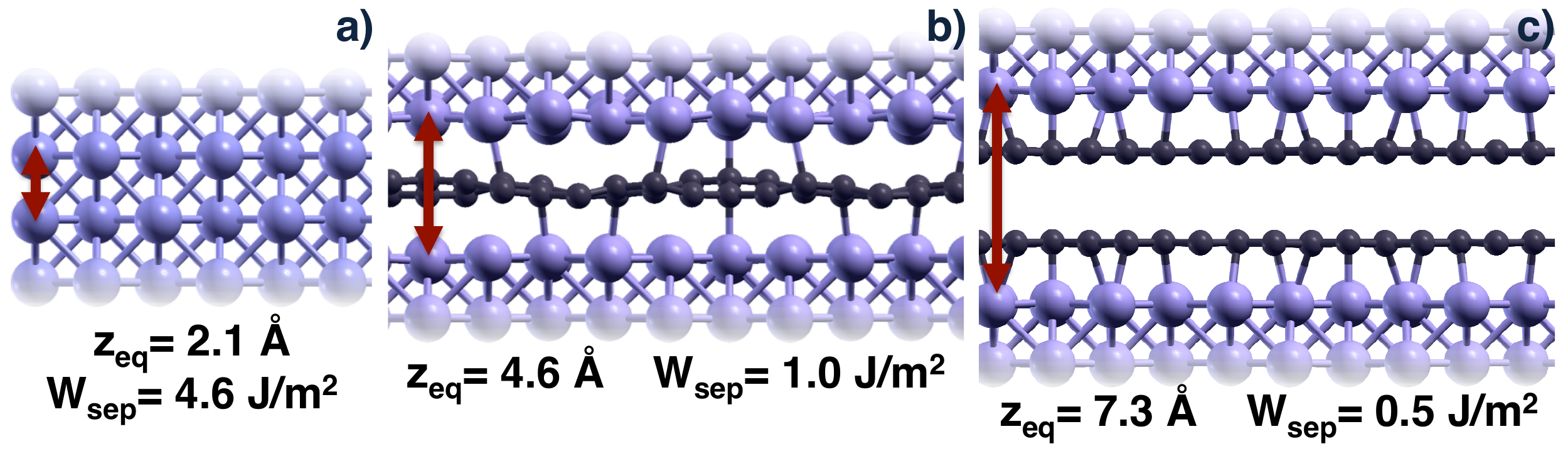}
\caption{Optimized structures of iron interfaces obtained by mating
two clean Fe(110) surfaces (a) a clean and a graphene-coated surface (b) and 
two surfaces coated by graphene (c). The equilibrium distance, $z_{eq}$, reached by the innermost
iron layers after the structural relaxation process (indicated by a red arrow) and the
work of adhesion, $W_{sep}$, are calculated
with the PBE approximation in (a) and within the PBE+vdW scheme in (b, c).}\label{interf}
\end{center}
\end{figure}

We, then, construct a second interface by mating a clean iron surface and a surface fully covered by graphene
(Fig.\ref{interf}b). It is very interesting to observe that the presence of an interfacial graphene
layer is able to decrease the interfacial adhesion by 78\%. The adhesion reduction reaches the 88\% when both
the surfaces in contact are coated by graphene, in this case the distance
between the mated iron surfaces becomes as large as $7.3 \; \text{\r{A}}$ (Fig.~\ref{interf}c). 
This last situation is representative of a condition
often occurring during pin-on-disc experiments, where part of the powder lubricant
is transferred from the substrate to the sliding pin. 

The above analysis reveals that the surface coverage by graphene is able to dramatically reduce the 
adhesion of iron surfaces and this impacts on the frictional properties, as shown in the following.
During sliding, the adhesion energy between two surfaces in contact varies with their relative lateral position and
this energy variation gives rise to frictional forces. In Fig.~\ref{pes} the PESes describing the work of separation 
as a function of the relative displacement of  two clean (a) and graphene-covered (b) iron surfaces are compared. 
The different symmetries of the two PESes reflect those of the Fe(110) and graphene lattices, respectively. But the most striking difference regards the PES corrugation,
which is more than thirty times higher in the PES of Fig.\ref{pes}a than
that of Fig.~\ref{pes}b. The MEP traced on the second PES 
is almost flat compared to that in the first one, as can be seen
from the energy profiles at the bottom of Fig.~\ref{pes}, which
cover energy-scales that differ by two orders of magnitude.
By derivative of the potential profiles we obtain the ideal shear strengths along the MEPs
that we assume as representative values for the shear strengths of the considered interfaces.
We obtain $\tau_{MEP}$ = 9.20 GPa for the clean iron interface, in agreement with previous 
DFT results derived by a different method based on the calculation of the Hellmann Feynman
stress tensor.\cite{ironshear_1,ironshear_2} 
The shear strength obtained for the iron interface 
covered by graphene is 98\% lower: $\tau_{MEP}$ = 0.17 GP. We, thus, conclude that iron coating by graphene
is extremely effective at reducing the intrinsic resistance to sliding of iron interfaces.

\begin{figure}[H]
 \begin{center}
\includegraphics[width=0.7\linewidth]{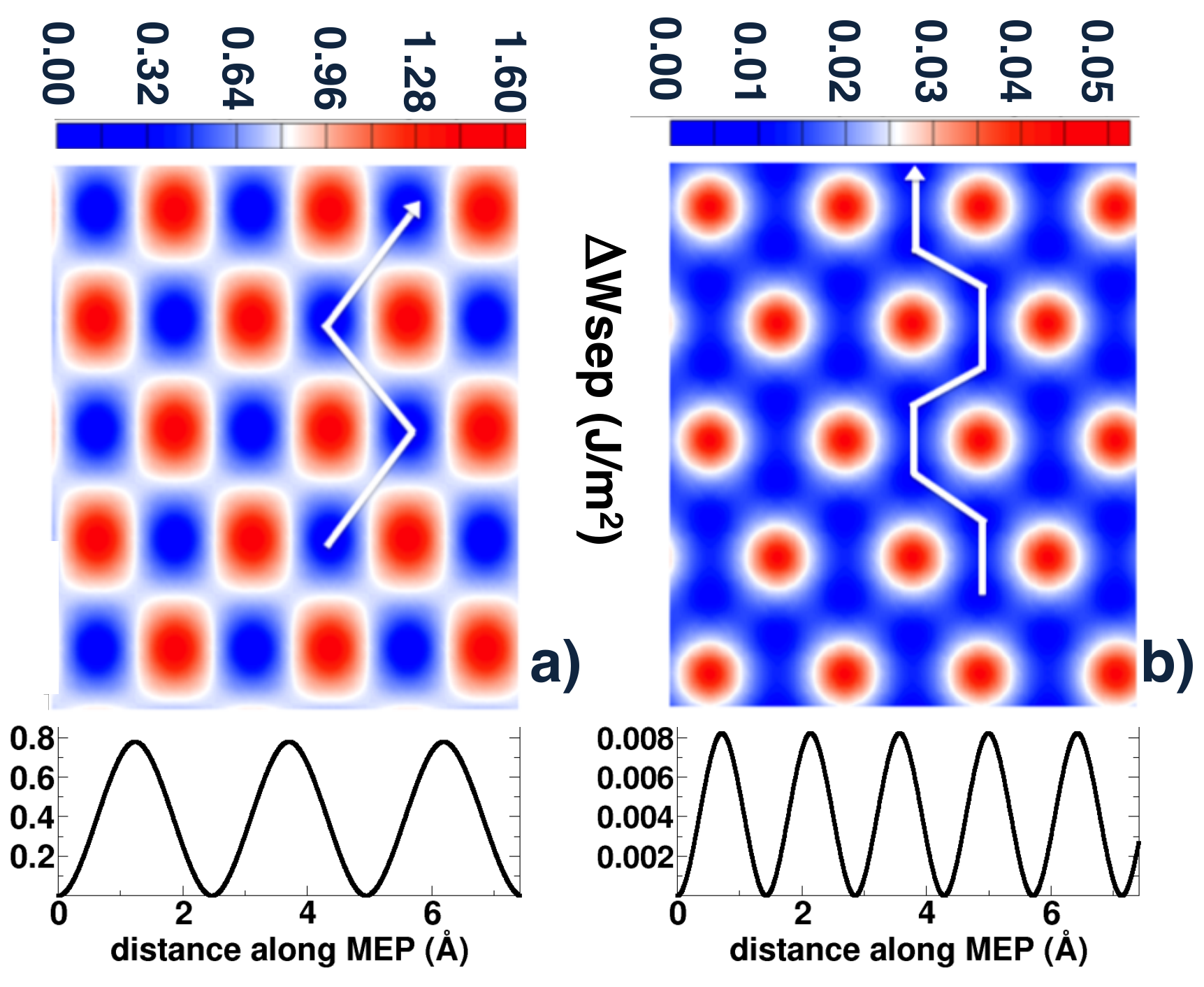}
  \caption{Potential energy surfaces (PESes) describing the variation of the interfacial work of separation (in J/m$^2$)
  as a function of the relative lateral position of two clean (a) graphene-coated (b) iron surfaces in contact. 
  The energy profiles along the MEPes (traced in white color on each PES) are reported at the bottom 
  of the figure. Different energy-scales are used.}  　\label{pes}
\end{center}
\end{figure}

\subsection{Correlation between the interfacial electronic charge and tribological properties}
To investigate the microscopic origin of the adhesion and friction
reduction provided by the graphene coating,
we first analyse the nature of the surface-surface interactions and then the electronic charge displacements occurring upon interface formation. In Fig.~\ref{ppes} the adhesion energy per unit area 
between two facing surfaces (corresponding to the opposite of $W_{sep}$) 
is reported as a function of their separation.
The comparison of the blue and black curves, corresponding respectively
to clean and graphene-coated iron interfaces,
reveals that the graphene coverage changes the nature of the surface-surface interaction from chemical to physical.
The black curve resembles, in fact, the curve describing the interaction between two isolated graphene layers (in red). Therefore, the graphene coating is able to passivate the iron surfaces very effectively, screening almost completely the metal-metal interaction at the interface. This may suggest that sliding micro-asperities fully covered by graphene may show similar friction anisotropy and ``superlubricity'' as observed in graphite.\cite{DienwiebelPRL,DienwiebelPRB,HodPRB}

\begin{figure}[H]
\begin{center}
\includegraphics[width=0.7\linewidth]{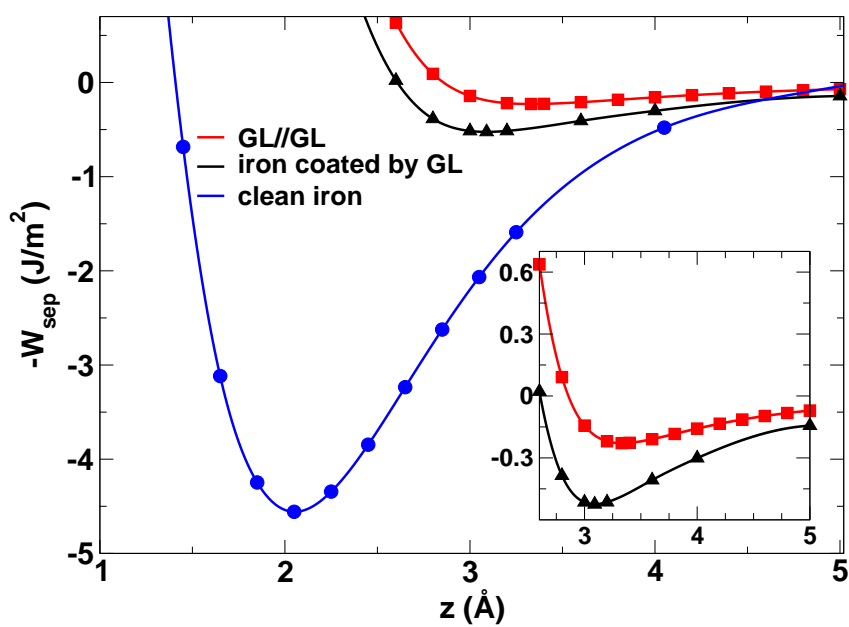}
\caption{Surface interaction energy as a function of separation
for the interfaces of Figs.~1a and 1c (blue and black lines). The binding energy curve for an isolated
graphene bilayer
is also displayed for comparison (red line).}\label{ppes}
\end{center}
\end{figure}

When two semi-infinite bulks are mated into an interface
an electronic charge redistribution occurs at their surfaces due 
the mutual interaction. Monitoring such charge redistributions 
provides fundamental insights into the microscopic origin of adhesion and  
frictional forces. The interfacial charge displacements can be obtained 
by first principles calculations as the difference between the electronic charge
of the interface and the sum of the electronic charges of the two isolated surfaces: 
$\Delta\rho({\bf r}) = \rho_{12}({\bf r})- (\rho_{1}({\bf r}) + \rho_{2}({\bf r}))$.
In Fig.\ref{charge} we plot the in-plane average $\Delta\bar{\rho}(z) = \Delta\rho(z)/A$, where $\Delta\rho(z)$ is obtained by 
in-plane integration of $\Delta\rho({\bf r})$ and $A$ is the surface area.
\begin{figure}[H]
\begin{center}
\includegraphics[width=0.7\linewidth]{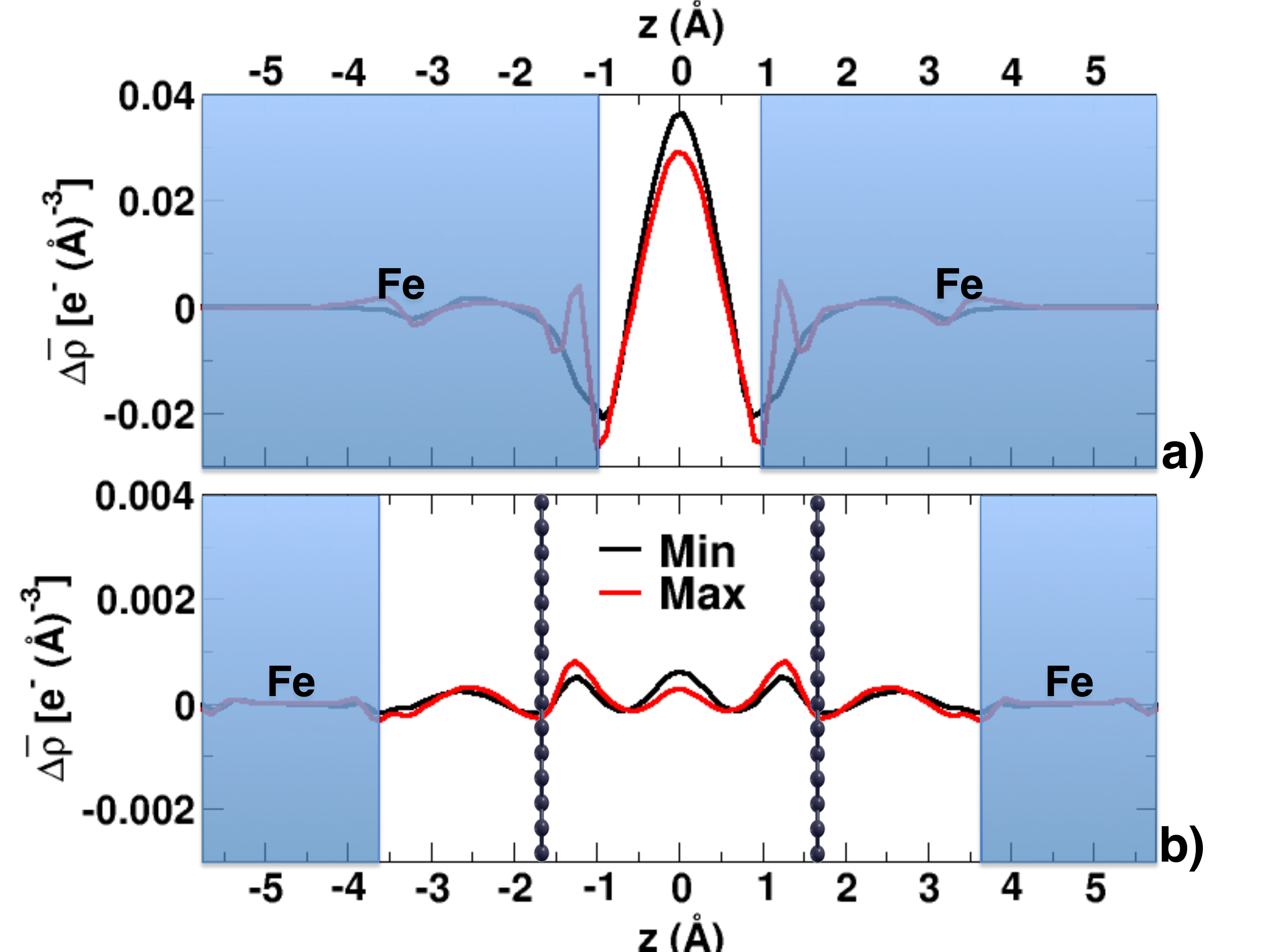}
\caption{Planar average of the electronic charge displacements 
occurring upon formation of a clean (a) and a graphene-passivated (b) iron interfaces. 
The white central areas correspond to the optimized interfacial spacings. Black and red colors
are used for the charge displacements calculated respectively for the most and less favorable
lateral positions of the two surfaces in contact.}\label{charge}
\end{center}
\end{figure}
By comparing the results obtained for the clean
iron interface (panel a) with that obtained for the 
iron interface coated by graphene (panel b), we can
observe two features: $i)$ When two surfaces are brought into contact,
a charge accumulation 
is set up at the middle of the interface
and the height of the central peak, $\Delta\bar{\rho}(z=0)$, is proportional to the magnitude of the interfacial adhesion,
$W_{sep}$ (note that two different scales are used).
$ii)$ The charge accumulation is higher for a relative lateral position
of the two surfaces corresponding to a PES minimum (black curves) than
for a PES maximum (red curves).
The difference between the maximum and minimum heights of the $\Delta\bar{\rho}(z=0)$ 
peaks is proportional to the potential corrugation, 
$\Delta W_{sep}$. 
As we describe in a forthcoming article, these two properties have general validity, $i.e.$ they
hold true in different materials. 
For the specific case we are considering, we can observe that the effect
of surface passivation by graphene is to
prevent the charge flow from the surfaces in contact towards
the interface, such inhibition of charge accumulation reduces the adhesion 
and friction of iron interfaces.

\subsection{Adhesion dependence on graphene coverage}
We conclude our investigation on the effects of graphene coating
on interfacial properties by analysing their dependence on graphene coverage.
We consider partial graphene coverage by 
modeling iron interfaces containing graphene ribbons.
The optimized structures shown in Fig.~\ref{coverage}
reveal that the ribbons remain
attached to the original substrate despite the presence of an iron counter-surface.
\begin{figure}[H] 
\begin{center}
\includegraphics[width=0.7\linewidth]{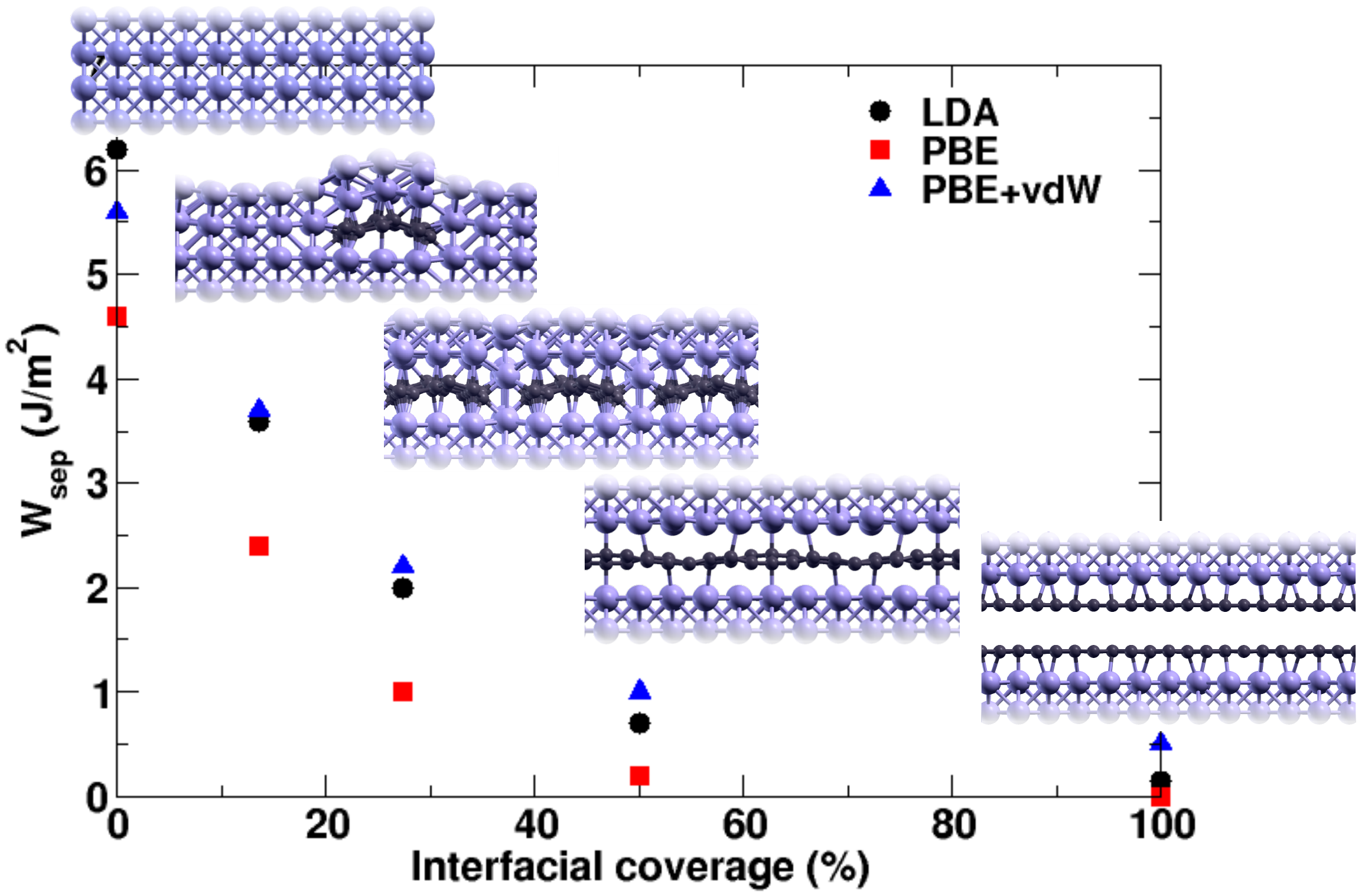}
\caption{Work of separation of iron interfaces as a function of graphene coverage.
Partial coating is obtained by considering adsorbed graphene ribbons.}\label{coverage}
\end{center}
\end{figure}
The works of separation, reported in Fig.~\ref{coverage}, reveal that the iron adhesion is considerably reduced
even in the case of partial surface coverage. The data trend, which is common to
all the three considered approximations (LDA, PBE and PBE+vdW), highlights a dramatic decrease of the interfacial adhesion with graphene coverage. This result suggests that the lubricating property of graphene, if ruled by the chemical interactions above described,  should be highly affected by the coverage.

\section{Conclusions}
\par In conclusion, we show that graphene chemisorbs on the iron surface
and its binding is enhanced by reactive dangling bonds,
as those present in pristine graphene flakes used in the colloidal lubricant.
By comparing the interaction potential between 
two clean and graphene-coated iron surfaces, we show that the surface passivation by graphene
is able to change the nature of the surface-surface interaction from chemical to physical
with a consequent drop of interfacial adhesion that can reach the 88\%
in case of full interfacial coverage. This adhesion reduction is accompanied by a 98\%
decrease of the ideal shear strength.
We show that a close correlation exists between the adhesion and the electronic charge 
accumulated at the interface.
Moreover, the evolution of the charge accumulation during sliding is proportional to the frictional forces.
The calculated adhesion and friction reductions are less consistent, but still present, in the case of 
partial interfacial coverage.\\
Our results are consistent with the Raman analysis performed after the tribological test,\cite{Erdemir_grafdry} which indicates that during the low-friction regime
graphene covers uniformly the wear track, while it is removed out from the track in the high-friction regime. 
We, thus, propose an explanation for the detrimental effect of load, which is alternative to that proposed by Klements \emph{et al.}\cite{Moseler} and consists in peeling off graphene from the surface.
This reduces the surface coverage by graphene and hence the passivation effect that provides lubricity.\\
We conclude by observing that the beneficial effect of surface passivation on the frictional 
properties of materials is not a new concept in tribology, 
as revealed for example by several studies on the tribochemistry of diamond/DLC in the presence
of passivating species like hydrogen and water molecules.\cite{carpick_prl,mio_jpcc}
Our results show that a lubricating effect of similar chemical nature, i.e. due to passivation
independently from graphene rupture, can be obtained by coating reactive metal surfaces by graphene. 

\section*{Acknowledgments}

We acknowledge the CINECA consortium for the availability of high
performance computing resources and support through the ISCRA-B TRIBOGMD
and ISCRA-C TIGra projects.

\section*{Appendix A. Supplementary data}

Supplementary data related to this article.

\section*{References}

\bibliography{fe-graphene}
\bibliographystyle{elsarticle-num}

\end{document}